# Precise manipulation of longitudinal dynamic self-assembly of particles in the viscoelastic fluid within a straight microchannel

Linbo Liu,[1, 4, a)] Haoyan Xu,[2, 4, a)] Haibo Xiu,[3] Nan Xiang,[1, b)] and Zhonghua Ni[1, b)]

*1. School of Mechanical Engineering, and Jiangsu Key Laboratory for Design and Manufacture of Micro-Nano Biomedical Instruments, Southeast University, Nanjing 211189, China*
*2. College of Control Science and Engineering, Zhejiang University, Hangzhou 310027, China.*
*3. College of Computer Science and Engineering, Zhejiang University, Hangzhou 310027, China.*
*4. John A. Paulson School of Engineering and Applied Sciences, Harvard University, Cambridge, Massachusetts 02138, United States.*
a) **Contributions:** *Linbo Liu and Haoyan Xu contributed equally to this work.*
b) *Author to whom correspondence should be addressed: nan.xiang@seu.edu.cn, nzh2003@seu.edu.cn*

The lack of a simple operable method for longitudinal dynamic self-assembly of particles in a microchannel is one of the main problems in applying this technology to a wide range of researches, such as biomedical engineering, material science, and computation. Herein, a viscoelasticity-induced trapping microfluidic system for flowing particles is proposed to increase the maneuverability of longitudinal dynamic self-assembly of particles and achieve real-time control of the interparticle spacings and the frequency of particles passing through the outlet. Two kinds of functional microstructures and a side-channel were designed to preprocessing the randomly distributed particles to make particles no aggregation and evenly distributed and realize real-time control of the particle volume concentration. Randomly distributed particles could be focused into a line and become equally spaced in the center axis of a straight microchannel under the balance of the elastic force and the viscoelasticity-induced effective repulsive force. Besides, a finite element method model is established to analyze the processes of particles flowing in each functional microstructure. Therefore, a step forward in this microfluidic technology can provide significant promotion for a wide range of researches.

In recent years, technologies for engineering two-phase microfluidic systems have catalyzed significant advances in several fields of science, such as material science,[1] biomedical engineering,[2] and computation.[3] Programmable and precision control of solid-liquid microfluidic systems, such as flowing particles in the microchannel, is of great significance for single-cell sequencing technologies,[4] fabrication of hydrogel,[5] and flow cytometry.[6,7] Besides, improving the uniformity of particles at the microscale, by using passive and contactless methods to counter the Poisson distribution can enable burgeoning fields such as tissue engineering and metamaterials synthesis. In tissue engineering, the arrangement of cells is a vital parameter for the design of printed tissues.[8] In metamaterials synthesis, complete control of the spatial composition of particles may allow tunable modulation of acoustic and optical properties of metamaterials.[9]

Recent progress in passive control of particle position in the cross-section of microchannel has taken advantage of various induced hydrodynamic effects in microfluidic systems. The widely used passive methods include inertial focusing,[10] hydrophoresis focusing,[11] and viscoelastic focusing.[12] Notably, the newly emerging viscoelastic focusing, which depends only on the intrinsic rheological properties of the fluids, has attracted increasing interests. Numerous studies have focused on the manipulation of the transverse position of flowing particles.[10] However, in addition to the transverse equilibrium state, there is a further equilibrium state in the flow direction. Flowing particles could self-assembled into equally spaced nodes under the single action of fluid flow, which can lead to a system for complete control of the flowing particles. Although studies up to now have provided simple descriptions and hypotheses on the mechanism of self-assembly in these two-phase solid-fluid systems, there still hard to employ this technology effectively in researches.[13]

Segré and Silberberg first observed the particle trains (strings of ordered particles) by using the inertial focusing principle, and explained that the formation of the ordered particles was due to particle-particle hydrodynamic interactions.[14] Di Carlo et al. achieved ordered particle phenomenon on multiple streamlines in a micro-scale channel with a square section.[15] Lee et al. proposed a hypothesis that short-chain particle train formation was due to the coupling of inertial forces and particle-particle hydrodynamic interactions. Furthermore, they achieved changes in the interparticle spacing, but it is irreversible.[16] In the above studies, particle trains flowed near the channel walls where the local shear gradient and particle rotation is large. Large shear gradient and rotation may result in fuzzy imaging and damaging the manipulated biological particles. Very recently, Del Giudice et al. achieved self-assembling particles into equally spaced nodes on the channel centerline, where both shear gradient and particle rotation are minimal.[17] Their work was conducted in non-Newtonian fluid, by adding polymer to aqueous suspensions to promote transversal migration of particles toward the centerline of a straight microchannel. However, the majority of existing studies employed inertial or viscoelastic fluids to achieve particle trains have only focused on finding parameters that matter and give simple theoretical hypotheses. Researchers still could not get an operable method to employ this technology in their studies.

In this letter, we report a simple operable method for self-assembling flowing particles into equally spaced nodes, a long consecutive particle train, in a square sectional straight microchannel. By using two kinds of functional microstructures combined with a side-channel for preprocessing flowing particles, particles could reach an equilibrium state in the flow direction and become equally spaced under the hydrodynamic particle-particle interactions. No filter or surfactant is required to pretreat the particle sample. Notably, the interparticle spacings and the frequency of particles passing through the outlet could be adjusted in real-time. A simple operable method is a crucial step forward for researchers to employ this technology in their studies and provides significant support for the applications in material science, biomedical engineering, and computation.

We employ the solution of hyaluronic acid (HA, molecular weight Mw = ~1.7 MDa, Sigma Aldrich) at mass concentrations of 0.3 wt % in DI water. HA 0.3 wt % is prepared by adding polymer powder to DI water at room temperature, and the solution is shaken to allow the dissolution of the polymer. Polystyrene particles (Sigma Aldrich) 20 μm in diameter (d) are added to HA solutions at volume concentrations $\varphi$ = 1.5 vol %. No surfactant is used to enhance dispersion, and no filter is used to remove potential aggregates.

Rheological measurements are implemented on a stress-controlled rheometer (Discovery HR-3) with a stainless steel cone and plate geometry (60 mm in diameter, 1°58`41`` angle). A solvent trap is employed to avoid fluid evaporation. The temperature is kept at T = 22°C. The fluid is using a high-precision syringe pump (PHD 2000, Harvard Apparatus, Holliston) to pumped at several volumetric flow rates Q (200-1800 μL/h). After flow stabilization, images are recorded using a high-speed camera (CLSM, Zeiss LSM510 META). The interparticle spacings $d_z$ is evaluated using a homemade Matlab program.

Precise manipulation of longitudinal dynamic self-assembly of particles is studied experimentally by suspending 20 μm polystyrene particles in an aqueous solution containing 0.3 wt % HA at particle loading of $\varphi$ = 1.5 vol %. As shown in Fig. 1(a), two kinds of functional microstructures and a side-channel were designed to preprocessing the randomly distributed particles to make particles no aggregation and evenly distributed and realize real-time control of the particle volume concentration. Then, a square-shaped straight microchannel with internal height H = 50 μm and total length L = 8 cm was employed to realize longitudinal dynamic self-assembly of particles. The particles go through three stages and two transitions in the long straight microchannel. Relatively evenly distributed particles could be focused into a line, reach the lateral equilibrium position. Then, the particles reach the longitudinal equilibrium state and become equally spaced under the hydrodynamic particle-particle interactions. The self-assembly states of particles were observed at a fixed distance from the starting position of the straight channel $L_s/H$ = 1600 ($L_s$ = 8 cm), the length of the observed portion is $L_{obs}$ = 80d = 1.6 mm, as shown in Fig. 1(b).

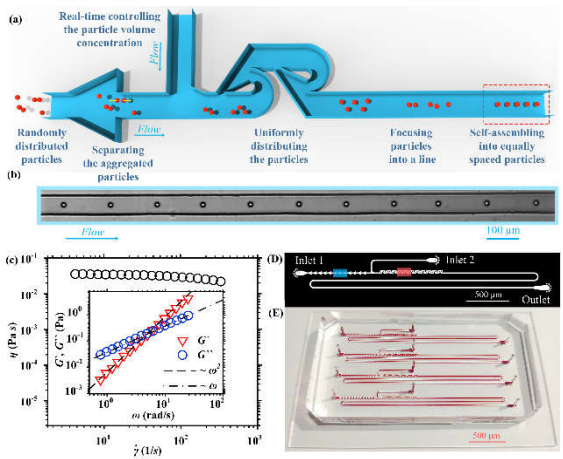

**Fig. 1.** (a) Schematic of the longitudinal dynamic self-assembling particle system. (b) Longitudinal dynamic self-assembly state of particles. (c) The shear viscosity $\eta$ as a function of the shear rate $\dot{\gamma}$ for hyaluronic acid (HA) at 0.3 wt % in DI water. The inset shows the elastic modulus $G`$ and the viscous modulus $G``$ as a function of the angular frequency $\omega$ for an imposed deformation $\dot{\gamma}$ = 5%. (d) The design of our microchannel. (e) The photo of our real Polydimethylsiloxane (PDMS) microfluidic chip.

For viscoelastic flows in square shape microchannels, there may simultaneously exist three types of hydrodynamic effects: elasticity effects, viscosity effects, and inertia effects. To quantitatively evaluate these effects, two dimensionless numbers were employed: Deborah number (De) and Reynolds number (Re). The Deborah number is defined to describe the ratio between the characteristic time λ of the fluid and the characteristic time $t_f$ of the flow. Besides, the Deborah number can also be regarded as a characteristic ratio between elastic and viscous forces in the flow field. The Reynolds number is defined to describe the magnitude ratio of inertial force to viscous force. The Deborah number and Reynolds number for a square shape channel are defined as follow:[17,18]

$$De = \frac{\lambda}{t_f} = \frac{\lambda Q}{H^3}, \quad (1)$$

$$Re = \frac{\rho U D_h}{\eta} = \frac{\rho Q}{\eta H}, \quad (2)$$

where Q is the volumetric flow rate, ρ is the fluid density, U is the average fluid velocity, $D_h$ is the hydraulic diameter of the channel, and η is the dynamic viscosity of fluids. As shown in Fig. 1(c), following a standard rheological procedure,[19] we obtain λ = 173 ms from the intersection of the dashed and the dashed-dotted lines fitting the data. The Deborah number and Reynolds number are De = 386 and Re = 0.026 respectively, which means the elastic force is the dominant force in this fluid system. The experiments were conducted in a PDMS chip, the microchannel pattern and photo of the PDMS chip are shown in Fig. 1(d-e).

After several preliminary experiments and iterative designs of the channel structure, we found that the initial state of the particles has a significant influence on whether the particle can complete the longitudinal dynamic self-assembly and, if so, the length of the channel required. Therefore, two kinds of functional microstructures and a side-channel were designed to preprocessing the randomly distributed particles to make particles no aggregation and evenly distributed and realize real-time control of the particle volume concentration. The first microstructure is an array of inverted triangular microchannels and straight microchannels, as shown in Fig. 2(a). The velocity of aggregated particles would be slowed in the inverted triangular microchannels until the former particle enters the straight microchannel. The velocity of the former particle tends to be increased, while the inverted triangular microchannel still slowed the latter particle so that there will be a separating force between the former particle and the latter particle. As a result, the aggregated particles get separated, which is also proved by the contour map of the flow velocity in the simulation result in Fig. 2(b). Then, the particle volume concentration of separated particles can be adjusted by regulating the flow ratio of the particle-free fluid in the side-channel and the particle fluid in the main channel. According to the conservation of mass, the interparticle spacings and the frequency of particles passing through the outlet after longitudinal dynamic self-assembly are mainly controlled by adjusting the particle volume concentration. Another functional microstructure is an array of swirl mixing channels.[20] After diluted by the particle-free fluid, the unevenly distributed particles will be thoroughly mixed and become evenly distributed, as shown in Fig. 2(c). The simulation result of flow velocity in Fig. 2(d) shows this microchannel has a good mixing ability. Evenly distributed particles have a higher possibility and require a shorter channel to be assembled into equally spaced nodes.

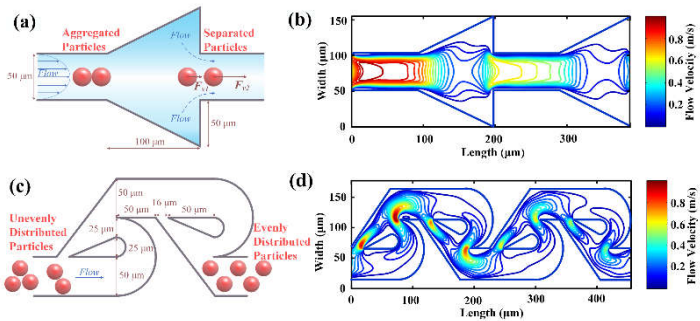

Fig. 2. (a) Schematic diagram of particles states and (b) contour maps of the flow velocity in an array of inverted triangular microchannels and straight microchannels. (c) Schematic diagram of particles states and (b) contour maps of the flow velocity in an array of swirl mixing channels.

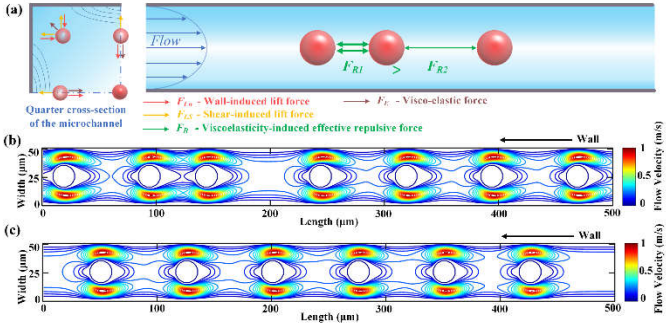

**Fig. 3.** (a) Left: Schematic illustration of the force mechanism of viscoelastic flows in a quarter cross-section of a square microchannel. Right: Asymmetric spacing around the particle generates an asymmetric viscoelasticity-induced effective repulsive force. (c-d) Contour maps of the flow velocity for (c) non-ordering and (d) ordering arrangement of the particles in square microchannels.

Fig. 3(a) shows that the forces on particles in our microfluidic system can be divided into three forces: visco-elastic force, wall-induced lift force, and shear-induced lift force.[21] As mentioned before, the elastic force dominant in our system so that particles will be focused on the center axis of the straight microchannel. Previous works highlighted that there are viscoelasticity-induced effective repulsive forces between neighboring particles and lead to ordering or pairing in particle trains.[17] This qualitative argument is consistent with our experimental phenomenons and simulation results in Fig. 3(b). In a long continuous particle train, the ideal interparticle spacing is a determinate value, say $s_1$, according to the conservation of mass and the particle volume concentration. If a small position perturbation of a particle generates an asymmetric spacing around the particle, the asymmetric repulsive forces ($F_{R1}$ and $F_{R2}$) will drive the train to recover its original spacing $s_1$, as shown in Fig. 3(c). As shown in Fig. 4(a), we can maintain the volumetric flow rate in side-channel on 500 μL/h and adjust the volumetric flow rate in the main channel from 250 μL/h to 1600 μL/h. Then we realize the real-time control of the interparticle spacings by adjusting the particle volume concentration from 0.52% to 1.15%. The statistics result in Fig. 4(b) shows that the interparticle spacings are nearly identical in each particle volume concentration in Fig. 4(a).

Conventional researches for longitudinal dynamic self-assembly of particles were mainly focused on its underlying mechanism and influence factors, which provides a good theoretical basis for us to develop a relatively maneuverable way for self-assembling particles into equally spaced nodes and real-time control the interparticle spacings. Our combined microstructures utilizing an array of inverted triangular microchannels and straight microchannels, swirl mixing channels, and a side-channel can separate aggregated particles and evenly distribute the particles and real-time control the particle volume concentration. Then, after self-assembly in an 8 cm long square-shaped straight microchannel, the particles become equally spaced at the particle volume concentration from 0.52% to 1.15%, also means realize the real-time control of the frequency of particle flowing pass the outlet from 280 Hz to 1600 Hz. A simple qualitative analysis using a simulation model demonstrates the hydrodynamic interactions and viscoelasticity-induced effective repulsive forces between the flowing particles in a long continuous train. Combined functional microstructures and a straight channel provide a path toward a maneuverable, inexpensive, and accurate way to real-time control of the longitudinal dynamic self-assembly of particles, which provide significant promotion for a wide range of researches, such as biomedical engineering, material science, and computation.

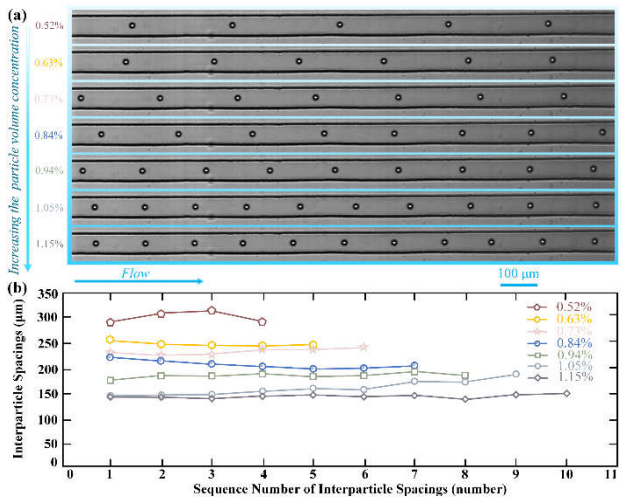

**Fig. 4.** (a) Snapshots of the high-speed movies and the reconstructed images show particles are self-assembled into equally spaced nodes in the flow direction. Each experiment process can be watched in supplementary movies S1-S7. The interparticle spacings can be controlled in real-time by adjusting the particle volume concentration. (b) The statistics of interparticle spacings, from (a). The interparticle spacings are nearly identical in each particle volume concentration.

See supplementary material for the device fabrication methods, numerical simulation, and supplementary experimental movie (Movies S1-S7).

This research work is supported by the National Natural Science Foundation of China (81727801，51875103, and 51775111), the Natural Science Foundation of Jiangsu Province (BK20190064), the Six Talent Peaks Project of Jiangsu Province (SWYY-005), the Zhishan Youth Scholar Program of SEU and the Jiangsu Graduate Innovative Research Program (KYCX17_0062).